\documentclass[final,3p,twocolumn,times,sort&compress,fleqn]{elsarticle}
\usepackage{graphicx}
\usepackage{amssymb}
\usepackage{amsmath}
\usepackage[small,bf]{caption2}
\usepackage[ruled]{algorithm2e}
\usepackage[bookmarksopen,bookmarksnumbered,colorlinks,linkcolor=blue,anchorcolor=blue,citecolor=green]{hyperref}
\usepackage[all]{hypcap}

\SetAlgorithmName{Algorithm}{Algorithm}{Algorithm}
\newtheorem{example}{Example}

\journal{Neurocomputing}

\begin{document}

\begin{frontmatter}


\title{Detecting Collusive Cliques in Futures Markets Based on Trading Behaviors from Real Data}
\author[dcse,shfe]{Junjie Wang}
\ead{wangjunjie@fudan.edu.cn}
\author[dcse,skliip]{Shuigeng Zhou\corref{col1}}
\ead{sgzhou@fudan.edu.cn}
\author[tongji]{Jihong Guan}
\ead{jhguan@tongji.edu.cn}

\cortext[col1]{Corresponding address: School of Computer Science,
Fudan University, 220 Handan Road, Shanghai 200433, China}
\address[dcse]{School of Computer Science, Fudan University, Shanghai 200433, China}
\address[shfe]{Shanghai Futures Exchange, Shanghai 200122, China}
\address[skliip]{Shanghai Key Lab of Intelligent Information Processing, Fudan University, Shanghai 200433, China}
\address[tongji]{Department of Computer Science \& Technology, Tongji University, Shanghai 201804, China}

\begin{abstract}
In financial markets, abnormal trading behaviors pose a serious
challenge to market surveillance and risk management. What is worse,
there is an increasing emergence of abnormal trading events that
some experienced traders constitute a collusive clique and
collaborate to manipulate some instruments, thus mislead other
investors by applying similar trading behaviors for maximizing their
personal benefits. In this paper, a method is proposed to detect the
hidden collusive cliques involved in an instrument of future markets
by first calculating the correlation coefficient between any two
eligible unified aggregated time series of signed order volume, and
then combining the connected components from multiple sparsified
weighted graphs constructed by using the correlation matrices where
each correlation coefficient is over a user-specified threshold.
Experiments conducted on real order data from the Shanghai Futures
Exchange show that the proposed method can effectively detect
suspect collusive cliques. A tool based on the proposed method has
been deployed in the exchange as a pilot application for futures
market surveillance and risk management.
\end{abstract}

\begin{keyword}
Futures markets \sep Financial trading behaviors \sep Collusive
cliques \sep Correlation coefficient \sep Weighted graph \sep
Unevenly-spaced time series. 
\end{keyword}

\end{frontmatter}

\section{Introduction}\label{sec:introduction}
In financial markets, trading behaviors roughly refer to
operations and actions conducted by individual investors to buy
and sell financial instruments through an exchange institute. Although
normal trading activities are dominating, abnormal market behaviors~(for example, price manipulation and circular trading)
happen now and then, especially in the emerging financial markets~\cite{Palshikar2000, Zhou2003, Walter2003, Khwaja2005, Palshikar2008}.
These abnormal behaviors not only impact market running mechanism
and pricing mechanism, but also threaten the safety of financial markets
and hurt the interests of righteous investors. What is worse, there is an
increasing emergence of abnormal trading events that for maximizing their personal benefits, some
traders constitute a collusive clique and collaborate with each other
to manipulate the movement of some instruments, thus mislead other
investors. Collusive trading
activities~\cite{Khwaja2005,Palshikar2008,Cao2010} are becoming a threatening and concealed type of financial market
manipulations. And discovering the hidden collusive cliques from numerous
market participants and massive trading data poses a tough challenge
to financial market surveillance and risk management, which thus has
attracted increasing attention of market regulators and researchers
in recent years. This is reasonable and natural when we consider this
issue under the situation that the world is still struggling from
the financial crisis.

The goal of this study is to detect collusive cliques in futures
markets based on similar trading behaviors of investors. Empirical
observation and analysis of trading operations of the market
participants can provide the clue to detecting the collusive cliques
in futures trading. The members of a clique are usually similar to each
other in trading behavior while different from the those outside the
clique. The similar trading behavior indicates that the members buy
or sell a certain instrument roughly at the same time point and even
their order volume is correlated. On the contrary, the trading
behaviors of ordinary (normal) investors who do not belong to any
collusive clique have little possibility of being correlated. Admittedly,
some ``clever" traders may attempt to take different operations for
counteracting collusive behaviors, which makes their activities
appear just as normal investors so that they can escape from being detected.
However, successful disguising needs not only high financial operation
skills on one hand but also extra cost on the other hand, which prevents
such collusive behaviors from happening popularly. This paper focuses on
detecting the first kind of collusive behaviors where individual investors
show (roughly) similar trading pattern, and leaves the problem of detecting
the second kind of collusive behaviors where individual investors must not
have similar trading fashion as future work.

In this paper, we propose an effective method to identify the
collusive cliques from numerous market participants. We first select
the dataset of real order records from the Shanghai Futures
Exchange\footnote{http://www.shfe.com.cn} by conducting a
comparative analysis on major information of futures trading
activities. Then, taking signed order volume as the characteristic
variable of futures trading activities, which can reliably reflect
the trading intentions of investors, we define a unified aggregated
time series to alleviate the disturbance caused by time difference
of trading event occurrences, and calculate the correlation
coefficient between any two eligible unified aggregated series.
Next, based on the correlation matrix of one trading day, a weighted
graph is constructed by using the edges whose weights are above a
predefined threshold. After that, the separate connected components
in the weighted graphs of multiple trading days are combined into an
integrated weighted graph where the weight of each edge is the sum
of its occurrences in different weighted graphs, and these edges
whose weights below a predefined threshold are given up. Finally,
the connected subgraphs in the integrated graph are taken as suspect
collusive cliques. Our method is mainly inspired by the empirical
observation and analysis on the real trading data, and we put the
first priority on the method's practicality in real applications of
market surveillance and risk management.

This paper is organized as follows: In Section~\ref{sec:related-work}
we provide a survey of some of the related work. The real dataset used in our study is
introduced in Section~\ref{sec:dataset}. Section~\ref{sec:methodology}
gives the detail of the proposed detection method and the concrete
algorithms. Experimental results are presented in Section~\ref{sec:experiments}.
Finally, Section~\ref{sec:conclusion} concludes the paper and highlights some
future works.

\section{Related Work}\label{sec:related-work}
To the best of our knowledge, there is no related work that detects
collusive cliques in futures markets based on similar trading behaviors as studied in
this paper. However, some works have dedicated to the problem of
detecting abnormal trading activities in financial markets from
different perspectives. For example, price manipulation, one of major
fraudulent trading activities, have been investigated by various methods,
including pattern recognition based approach~\cite{Palshikar2000},
behavioral statistic model~\cite{Zhou2003, Hansen2004, Khwaja2005,
Aggarwal2006}, rational expectation theory of corners
~\cite{Allen2006} and domain driven data mining~\cite{Ou2008}.

As an emerging kind of abnormal activities in financial markets,
collusive activities among investors recently have been
investigated from different aspects for explaining market manipulation.
For distinguishing the irregular trading patterns from the regular
trading operations, Franke et al~\cite{Franke2007} developed
detection approaches based on spectral clustering method. They
generated a trader network to represent the trading behaviors of
traders and thus characterized the market. If the actual market
behaviors deviate from the allowed trading behaviors in the market,
then irregularities are reported. However, this study was conducted
on an experimental stock market. Palshikar et al~\cite{Palshikar2008}
proposed a graph clustering algorithm for detecting a set of
collusive traders who have heavier trading among
themselves compared to their trading with the other traders. They
constructed stock flow graph with synthetic trading data to
represent the trading relationships between traders, and applied the
graph clustering method to find collusive traders. Cao et al~\cite{Cao2010} argued that market
manipulation derives from the activities of a group of hidden
manipulators who collaborate with each other to manipulate three
trading sequences: buy-orders, sell-orders and trades, through
carefully arranging their prices, volumes and time. They proposed a
a coupled Hidden Markov Models(HMM)-based approach to describing the
interactive behaviors among group members, and further to detecting
abnormal manipulative trading behaviors on orderbook-level stock
data.

Comparing with these works above, our study in this paper has three distinct
features as follows: first, our work addresses
collusive clique detection in futures markets, while the existing
works all studied irregularity discovery in stock markets. Although
both futures and stock are financial products, their trading
mechanisms are quite different. Second, we build weighted graphs to
characterize the interactions among the investors based on their
trading behaviors, which is different from the existing works that
also used graph based approaches. Last but not least, our method is
inspired by and evaluated with real order data of futures trading in
the Shanghai Futures Exchange. However, most existing works (not
including Cao et al~\cite{Cao2010}) evaluated their methods by
synthetic data.

In fact, the detection of collusive behaviors has also studied in other
fields, including online auction systems
~\cite{Trevathan2007a, Trevathan2007b}, online recommender systems
~\cite{Lam2004,SuXF2005,Chirita2005,ZhangS2006}, online reputation systems
~\cite{Zhangh2004, Wangjc2008, Liuyh2008} and P2P file sharing networks
~\cite{Feldman2004,Lianq2007}. Solutions of these systems are
effective in the respective scenarios, but none of them could be applied
to the detection of collusive cliques in the futures markets for
three reasons. First, trading activities in the futures markets are
complicated. For example, every investor can continuously open or close
long/short positions in any futures contract. Second, there are
hundreds of thousands of order recodes submitted to futures trading
systems in one typical trading day. Such large scale data sets
could not appear in most online auction systems or reputation systems.
Last but not least, behavioral ratings and interaction between two colluders
are not the ideal description of collusion behaviors for high
frequent order sequences in the futures trading systems.

In our study, one key technique for detecting of collusive cliques
is the measure of similarity between a pair of unevenly-spaced time
series. The similarity of time series has been measured by various
metrics, including \emph{Euclidean distance} (ED)~\cite{Agarwal1993,
Keogh2002} and more sophisticated metrics, such as \emph{Dynamic
Time Warping}~(DTW) \cite{Berndt1994, Keogh2000}, \emph{Edit
distance with Real Penalty} (ERP) \cite{Chen2004}, distance based on
\emph{Longest Common Subsequence} (LCSS)~\cite{Vlachos2002},
\emph{Edit Distance on Real sequences} (EDR)~\cite{Chen2005},
\emph{Spatial Assembling Distance} (SpADe)~\cite{Chen2007} and
\emph{Sequence Weighted ALignmEnt} model (SWALE)~\cite{Morse2007}.
These representation and distance measures mentioned above have been
comprehensively evaluated by comparative experiments in~\cite{DingH2008}
for querying and mining of time series database.
These methods try to identify matching elements between time series.
However, the trading behavior similarity of two investors in
collusive clique detection is characterized by the conformity and
correlation between the pair of corresponding time series of signed
volume, which emphasizes the shape similarity instead of magnitude
similarity. Time stamp of trading activity is a
principal characteristic to evaluate the similarity between two time
series of signed volume. Two time series even with the same shape
happening in different time periods could not been considered to be
similar. For these reasons, correlation measurement is more
appropriate for our study.

\section{The Dataset}\label{sec:dataset}
Data is the key to data mining. Understanding the data is crucial to
the design of data mining algorithms. In this section, we will
introduce the dataset used in this study for detecting collusive
cliques in futures markets.

In futures trading, there are different types of data, such as order
records, trade results and position changes, which can provide clue
to describing the trading behavior of a market participant. An
\emph{order} is an instruction to buy or sell instruments, submitted
by an investor to the electronic trading platform of the exchange
institute. The order record indicates the trading intention of the
investor to buy or sell how much volume of a specific instrument at
the price of the moment. The eligible orders from buyers and sellers
are matched according to a certain rule via the electronic trading
platform, and trade reports are sequentially generated for the
investors. Both the dealing prices and the trade volumes of the
transactions are derived from the corresponding orders and are
dependent on the current market situation such as the last prices
and the order volumes of counterparts. The trading results will lead
to position changes of the involved investor. Therefore, both
trading results and position changes are the derivative consequences
of order records, they can only partly represent the investors'
intentions. However, order information can properly characterize the
investors' trading behaviors.

The dataset used in our investigation is entirely from the real
order series of the Shanghai Futures Exchange, which is the largest one in
China's domestic futures market and has considerable impact on the
global derivative market. Currently, the electronic trading platform
of the exchange institute receives only limit orders submitted by
the investors. There are hundreds of thousands of order records from
market investors in one typical trading day, which is comprised of
the open call auction (8:55 - 8:59) and four continuous auction
sessions (9:00-10:15, 10:30-11:30, 13:30-14:10 and 14:20-15:00). We
collect a representative order dataset that cover three active
futures contracts, including copper, fuel oil and natural rubber in
the nine trading days from Sep 16, 2008 to Sep 26, 2008. The dataset
contains 1,893,519 order records and involves 66,861 market
participants.
\begin{table}[tb]
    \captionstyle{indent}
    \caption{The statistic information of the order dataset of three futures
    contracts, including copper, fuel oil and natural rubber in the nine trading days from Sep 16, 2008 to Sep 26, 2008.}
    \label{tab:odssi}
    \centering
    \begin{tabular*}{0.48\textwidth}{@{\extracolsep{\fill}}ccc }
        \hline
        Futures&Number of orders&Number of investors\\
        \hline
        \itshape copper         & 441,104 & 19,414 \\
        \itshape fuel oil       & 650,079 & 22,537 \\
        \itshape natural rubber & 802,336 & 24,910 \\
        \hline
    \end{tabular*}
\end{table}

The statistic information of the order records of the three futures
contracts is given in~\autoref{tab:odssi}. A limit order record
includes a virtual ID representing the investor, bid/ask indicator,
order price and volume. All other sensitive information is filtered
out for privacy preservation reason.

\section{Methodology}
\label{sec:methodology} In this section, we will describe the detail
of the method employed to detect collusive cliques by calculating
correlation coefficients between trading series and constructing
weighted graphs from the correlation coefficient matrices. The
algorithms for implementing the method are also given.

\subsection{Selection of the Target Variable}
A \emph{limit order} refers to an order submitted by an investor to
buy or sell an instrument at a specific price (rather than a market
price), thus it contains the fundamental information such as bid/ask
indicator, order price and order size. For these fields of a limit
order record, which one can be used as a representative data item to
describe the trading intention of an investor? To answer this
question, let us first check these fields in detail.

As a piece of crucial information, the \emph{bid/ask indicator}
indicates whether the order is a buy limit order or a sell limit
order and whether the investor wants to own or to abandon the asset.
The \emph{order price} is a specific price at which the investor
hopes the order will be filled. Generally, the price that is close
to the latest trade price of the market will be immediately filled,
and the prices of orders submitted during a short period are almost
the same. Consequently, the price that is dependent on the market
situation does not distinguish the investors' intentions. The
\emph{order volume} reflects the amount of asset that a investor
intends to buy or sell.

Based on the preceding analysis of different fields in limit order
records, we decide to combine the \emph{order volume} and the
\emph{bid/ask indicator} into a \emph{signed order volume} as the
proper representation of a participant's trading intention. A signed
order volume sequence denotes the \emph{bid indicator} with positive
sign for buy order, and denotes the \emph{ask indicator} with
negative sign for sell order. That is, in a signed order volume
sequence, the volume of a buy order is positive while the volume of
a seller order is negative. By using signed order volume to describe
a trading event of a certain investor at the moment of submitting
her/his order, a discrete event sequence over a period of trading
time can naturally characterizes the trading behavior of an
investor.

For a futures investor, we denote by $v(t_i)$ the signed order
volume of an order submitted by her/him at time $t_i$ ($i$=1, 2,
$\cdots$, $N$). $N$ is the length of the sequence $\{t_i\}$, which
may be different for different investors, and the time points of the
sequences for different investors can also be different. Thus, the
time series $\{v(t_i)\}$ of the signed order volume is an
unevenly-spaced event sequence.

\begin{example}\autoref{tab:ssc} illustrates the
construction of two signed order volume sequences from the limit
order sequences of two investors \#1 and \#2. In the table, the
first five columns are the information fields of limit orders, and
the last column represents the signed order volume. The
\emph{timestamp} of limit order is in the format using the colon as
the separation character.
\end{example}

\begin{table*}[t]
    \captionstyle{indent}
    \caption{The limit order sequences of two investors and the corresponding signed order volume sequences.
    The first five columns are the information fields of limit orders. The last column represents the signed order volume whose positive sign means buy order and negative sign means sell order.
    }
    \label{tab:ssc}
    \vspace{2mm}
    \centering
    \begin{tabular*}{\textwidth}{@{\extracolsep{\fill}} cccccc }
        \hline
        Investor & Timestamp & Indicator & Price & Volume & Signed volume\\
        \hline
        1 & 09:00:30 & Buy  & 3211 & 2 &  2 \\
        1 & 09:03:06 & Sell & 3216 & 2 & -2 \\
        1 & 09:03:12 & Sell & 3214 & 1 & -1 \\
        1 & 09:08:02 & Sell & 3206 & 2 & -2 \\
        1 & 09:08:26 & Buy  & 3204 & 6 &  6 \\
        1 & 09:10:28 & Sell & 3205 & 3 & -3 \\
        \hline
        2 & 09:00:40 & Buy & 3211 &  3 &  3 \\
        2 & 09:03:04 & Sell& 3216 &  4 & -4 \\
        2 & 09:03:10 & Buy & 3214 &  2 &  2 \\
        2 & 09:08:05 & Sell& 3206 &  3 & -3 \\
        2 & 09:08:30 & Buy & 3204 & 10 & 10 \\
        2 & 09:12:02 & Buy & 3201 &  2 &  2 \\
        \hline
    \end{tabular*}
\end{table*}

\subsection{Aggregated Time Series}
The event series of signed order volume above is not appropriate to
calculate the behavior similarity of different investors due to two
reasons as follow. On one hand, even though two investors belonging
to a clique desire to apply the same order strategy, their
operations can not be accurately synchronous in practice, usually
there exists a little lag for some reasons (e.g., network speed or
the queuing policy of the exchange). On the other hand, the active
speculators such as day traders always issue a large number of order
records, their long event sequences make the computation of behavior
similarity more complex. Therefore, here we introduce an
\emph{aggregated sequence} to replace the original signed order
volume sequence to represent the behavior of an investor.

We specify the size $\delta_t$ of a time window. Given a signed
order volume sequence, we split the sequence from its starting
timestamp into a series of consecutive windows (or segments) of
length $\delta_t$, each of which is labeled by its time index whose
value is an positive integer starting from 0. That is, the first
window is labeled by 0, the second one is by 1, and so on. For the
\emph{i}-th window, its time index is denoted by $s_i$, and it
covers the scope of time $[s_i \delta_t, (s_i+1) \delta_t)$. We
aggregate the signed volumes of different orders happening within
each window into a single value. Concretely, for the \emph{i}-th
window, the aggregation value $V(s_i)$ is the sum of all signed
volumes of orders happening in $[s_i \delta_t, (s_i+1) \delta_t)$.
Formally,
\begin{equation} \label{eq:ATS}
V(s_i)={\sum_{s_i \delta_t \le t_j < (s_i+1) \delta_t}{v(t_j)}}.
\end{equation}
Above, $v(t_j)$ is the signed volume of an order happening at the
timestamp $t_j$ in the signed order volume sequence under study.
Thus, a signed order volume sequence can be transformed to an
aggregated sequence, denoted by \{$($$s_i, V(s_i)$$)$\}~($s_i$=0, 1,
2, $\cdots$). Furthermore, we discard any aggregated point $s_i$
whose aggregated value $V(s_i)$=0, then get the final
\emph{aggregated signed order volume sequence}, which is an
aggregated time series.

\begin{example}\label{exp:aggregated-time-series}
For the data in \autoref{tab:ssc}, let $\delta_t$=60 seconds and
the starting timestamp of the order series is 09:00:00, we can get
the time index sequences of the aggregated sequences for the two
investors \#1 and \#2 are \{0, 3, 8, 10\} and \{0, 3, 8, 12\} ,
respectively. And the two signed order volume sequences are
aggregated into two aggregated signed order volume sequences as
follows: \{(0, 2), (3, -3), (8, 4), (10, -3)\} and \{(0, 3), (3,
-2), (8, 7), (12, 2)\}.
\end{example}

In practice, there will be no aggregated data in a certain time span
if there is no order event occurring at all during that period, thus
the aggregated time series is unevenly spaced. The time window size
$\delta_t$ determines the granularity of aggregation and the length
of the aggregated series. By enlarging the window size, buying and
selling volumes within a window may counteract, which thus makes the
aggregated value of that window smaller and consequently degrades
the calculation result. Therefore, a reasonable time window size is
critical to the calculation of behavior correlation coefficient.

Furthermore, the collusive investors tend to frequently place orders
to influence the market, they easily become the active traders in
the market. Consequently, the investors with few orders will very
possibly be excluded from the detected potential collusive cliques
because they will not be highly correlated with these investors who
have more orders. To reduce the unnecessary computation and thus
boost efficiency, we filter out some investors who have few orders
before correlation coefficient computation. Concretely, we compare
the length of each aggregated time series with an empirical
threshold~($\delta_L$), and only these with a length no shorter than
the threshold are kept for further processing. We call these
aggregated time series \emph{eligible aggregated signed order volume
series}. So only the eligible aggregated signed order volume series
will be used for correlation coefficient computation and potential
collusive cliques detection.

\subsection{Unified Aggregated Time Series and Correlation Measure}
The trading behavior similarity between two investors is evaluated
by the strength of association between the corresponding aggregated
time series, which is commonly measured by correlation
coefficient~\cite{Rodgers1988, Stigler1989}. In statistics,
correlation coefficient is used as an indicator of the degree to
which an event or phenomenon is associated with, related to, or can
be predicted from another, as well as a strength measure of linear
relationship between two variables. It has been widely applied in
the financial area. We adopts correlation coefficient as the
similarity measure of two investors' trading behaviors.

Let us consider two investors $A$ and $B$, and their aggregated time
series are denoted by $V_A$ and $V_B$, respectively. Both aggregated
time series are unevenly spaced and discrete, and their time indices
$s_i^A$ and $s_i^B$ with the same subscript $i$ are not necessary
the same, thus the methods for evenly-spaced time series are not
applicable here. We merge their time index sets $s^A$ and $s^B$ into
a unified time index set $s$, i.e., $s=s^A \cup s^B$. Based on the
unified time index set, we define the \emph{unified aggregated time
series} of signed order volume of investor $A$ as follows:
\begin{equation}\label{eq:ASR}
U_A(s_k)=\left\{
\begin{array}{l l}
V_A(s_k), & s_k \in s^A \\
0. & otherwise
\end{array}
\right.$$
\end{equation}
Similar definition is applied to the aggregated time series $V_B$ of
investor $B$, and we get the \emph{unified aggregated time series}
$U_B$ of investor $B$.

\begin{example}\label{exp:unified-aggregated-time-series}
Following Example~\ref{exp:aggregated-time-series}, we go further to
compute the unified aggregated time series of investor \#1 and \#2.
The unified time index set is $\{0, 3, 8, 10\} \cup \{0, 3, 8,
12\}=\{0, 3, 8, 10, 12\}$. According to ~\autoref{eq:ASR}, we have
the unified aggregated time series $U_1$ and $U_2$ as follows: $U_1$
= \{(0, 2), (3, -3), (8, 4), (10, -3), (12, 0)\}, $U_2$ = \{(0, 3),
(3, -2), (8, 7), (10, 0), (12, 2)\}.
\end{example}

The correlation coefficient is evaluated between unified aggregated
time series. For $U_A$ and $U_B$, their correlation coefficient
$r_{AB}$ is defined as follows:
\begin{equation}\label{eq:cc}
    r_{AB}={{<U_A U_B>-<U_A><U_B>} \over {\sqrt{(<U_A^2>-<U_A>^2)(<U_B^2>-<U_B>^2)}}}
\end{equation}
where the angular brackets $<\cdots>$ represents the average over
all the aggregated events (or points) in the series. The correlation
coefficient $r$ is between -1 and 1. A positive $r$ value indicates
the existence of positive correlation, while a negative $r$ value
implies negative correlation. A zero $r$ means no correlation and
the two time series are independent from each other. For collusive
clique detection, negative correlation is little significant because
it means the trading behaviors of two investors are almost opposite.
In fact, only positive correlation is of significance for collusive
cliques detection.

\begin{example}
Following Example~\autoref{exp:unified-aggregated-time-series},
according to ~\autoref{eq:cc}, the correlation coefficient between
the two unified aggregated time series $U_1$ and $U_2$ is 0.956,
which means that the trading behaviors of the two investors are
strongly positive correlated.
\end{example}

When considering $N$ investors, the correlation coefficients between
any two investors $i$, $j$ build a correlation matrix $R$, which is
an $N \times N$ matrix where the entry $r_{ij}$ indicates the
correlation coefficient between two unified aggregated time series
$U_i$ and $U_j$. The correlation matrix is symmetric because the
correlation between $U_i$ and $U_j$ is the same as the correlation
between $U_j$ and $U_i$. The diagonal elements in the matrix are the
self-correlation coefficients of all unified aggregated time series,
and the values are 1.

\subsection{Discovery of collusive cliques}
With the correlation coefficient matrix, we construct a weighted
graph in which a node represents an investor in the market, and an
edge is added to connect two nodes if the correlation coefficient
between the two nodes' corresponding unified aggregated time series
is larger than a user-specified threshold ($\delta_w$). Note that
the weighted graph constructed does not contain loop edges and there
is no more than one edge between any two nodes, and the weight of
each edge is the correlation coefficient. The resulting graph is not
necessary a connected graph, very possibly it may includes some
isolated nodes and some connected components (subgraphs). An
isolated node has no link to any other nodes, and a connected
component may be a complete graph, which means that all nodes in the
component are highly similar to each other but weakly similar to the
other nodes outside the component. Obviously, the connected
components conform to the criterion of potential collusive cliques.
Certainly, the value of correlation coefficient threshold $\delta_w$
will surely influence the number of resulting connected components.
As the threshold increases, the number of resulting connected
components will reduce, and the detected result will be more
reliable but some suspect traders may be neglected. On the contrary,
when decreasing the threshold, the number of false collusive cliques
(noise) will rise, which will degrade the detection precision.
Therefore, a proper threshold $\delta_w$ is of substantial
importance to guarantee the detection performance.

It is always the case that we do not know how many collusive cliques
exist and who belongs to which clique in the market. Fortunately,
some practical observation can help us to make the decision. That
is, a collusive clique will conduct cooperative and abnormal actions
repeatably. So if a doubtful clique happens in multiple trading
days, it is reasonable to believe that it is a suspect collusive
clique. Thus we consider multiple continuous trading days, and
construct one weighted graph for each trading day, then combine the
connected components in the daily graphs into an integrated weighted
graph in which the weight of each edge is the sum of its occurrences
in different weighted graphs. Finally, the connected subgraphs in
the integrated graph are output as suspect collusive cliques by
eliminating the isolated nodes and the edges whose weights are below
a predefined threshold~($\delta_f$).

\subsection{The algorithms}
Our method of collusive clique detection by similar trading behavior
analysis mainly consists of two stages:
\begin{itemize}
\item Computing the unified aggregated time series of signed order volume for each investor,
and calculating the correlation coefficient matrix based on all
eligible unified aggregated time series, and \item Identifying
suspect collusive cliques by combining the connected components in
the weighted graphs of multiple continuous trading days into an
integrated weighted graph.
\end{itemize}

We develop two algorithms to implement the tasks of the two stages
above. They are outlined in \autoref{algo:cccm} and
\autoref{algo:dccc}.
\begin{algorithm}[!h]
    \SetAlgoLined
    \caption{Calculating correlation coefficient matrix}
    \label{algo:cccm}
    \KwIn{Order record set $D$ of one futures contract in one trading day, time window size $\delta_t$, length threshold $\delta_L$ of aggregated time series}
    \KwOut{Correlation coefficient matrix $R$}

    $T := \emptyset$\;
    \For{each investor $p$}
    {
        Extract time series $v_p$ of signed order volume from $D$\;
        Aggregate $v_p$ by summing up signed order volumes in each time window $s$ of size $\delta_t$. The aggregated time series is denoted as  $V_p$\;
        \If{$|V_p| >= \delta_L$}
        {
            Add $V_p$ to $T$\;
        }
    }
    \For{each $V_i, V_j \in T, i \neq j$}
    {
        Merge the two time index sequences $s^i$ and $s^j$ into an unified time index sequence $s$ with $s=sort(s^i \cup s^j)$\;
        Unify $V_i$ based on $s$ into $U_i$ by
        $\{U_i(s_k)\}=\{V_i(s_k)|s_k \in s^i \} \cup \{0|s_k \in s, s_k \notin s^i\}$, Unify $V_j$ into $U_j$ in the same way\;
        Calculate correlation coefficient $r_{ij}$ between $U_i$ and $U_j$ according to the following formula:\
        $r_{ij}={{<U_i U_j>-<U_i><U_j>} \over {\sqrt{<U_i^2-<U_i>^2><U_j^2-<U_j>^2>}}}$\;
    }
    Output $R$\;
\end{algorithm}

\autoref{algo:cccm} aggregates the signed order volumes of a market
investor in every time window of a trading day to a single value,
and filters out the short aggregated time series, and then
calculates the correlation coefficient between any two unified
aggregated time series. This algorithm's input is the preprocessed
order records set of one futures contract in a single trading day
and each order record includes the investor's virtual ID, signed
order volume and a second-based timestamp converted from the time
format that uses the colon as separation character.
\begin{algorithm}[h]
    \SetAlgoLined
    \caption{Detecting collusive cliques}
    \label{algo:dccc}
    \KwIn{Correlation coefficient matrix set $\{R^d\}$ for one futures contract in multiple trading days,
    threshold $\delta_w$ for constructing weighted graphs, edge weight threshold $\delta_f$ for the integrated weighted graph.}
    \KwOut{Candidate collusive cliques}

    \For{each correlation matrix $R^d$}
    {
        Construct a simple weighted graph using $R^d$ as the adjacent matrix, in which an edge exists if its weight is greater than $\delta_w$\;
        Obtain the connected components set $S^d$ of the graph\;
    }
    Merge $\{S^d\}$ into an integrated weighted graph $G$, in which the weight of each edge is the sum of its occurrences in $\{S^d\}$\;
    Eliminate the edges whose weights are below $\delta_f$\;
    The connected subgraphs in $G$ are output as potential collusive cliques\;
\end{algorithm}

With the correlation coefficient matrix, \autoref{algo:dccc} is
developed to detect collusive cliques. It first constructs one
weighted graph for each of trading day, and then merges the
connected components of the daily graphs into an integrated weighted
graph. For each connected component in the integrated weighted
graph, if the weights of all its edges are no less than the
threshold $\delta_f$, then the connected component is output as a
suspect collusive clique.

\section{Experiments and Discussions}\label{sec:experiments}

In this section, we will present the experimental results with real
order data of three futures contracts from the Shanghai Futures
Exchange. The experimental results confirm the effectiveness of the
proposed method in detecting collusive cliques.

\subsection{The Effect of Time Window Size $\delta_t$}
\begin{figure}[tb]
    \centering
    \includegraphics[width=0.497\textwidth]{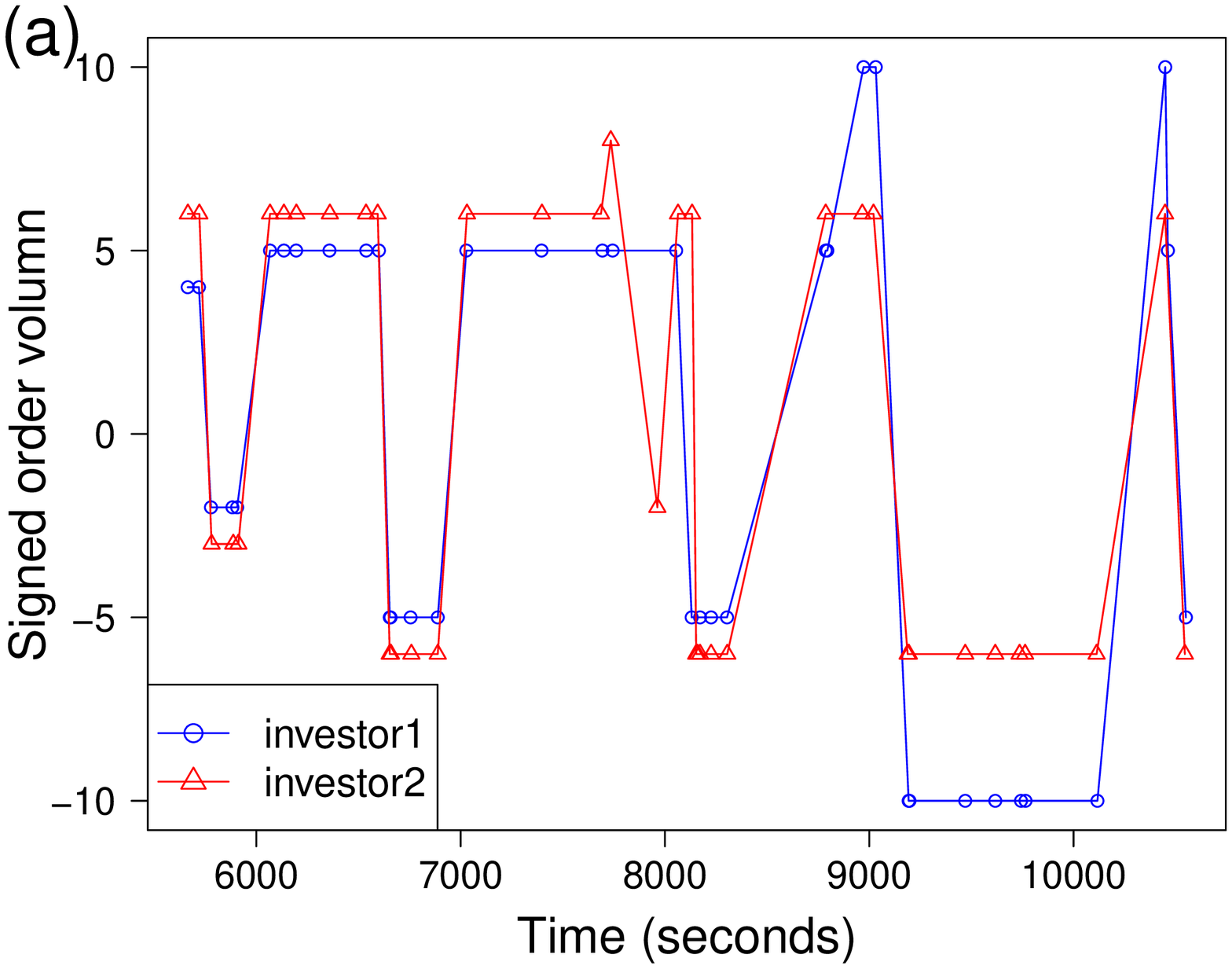}
    \includegraphics[width=0.497\textwidth]{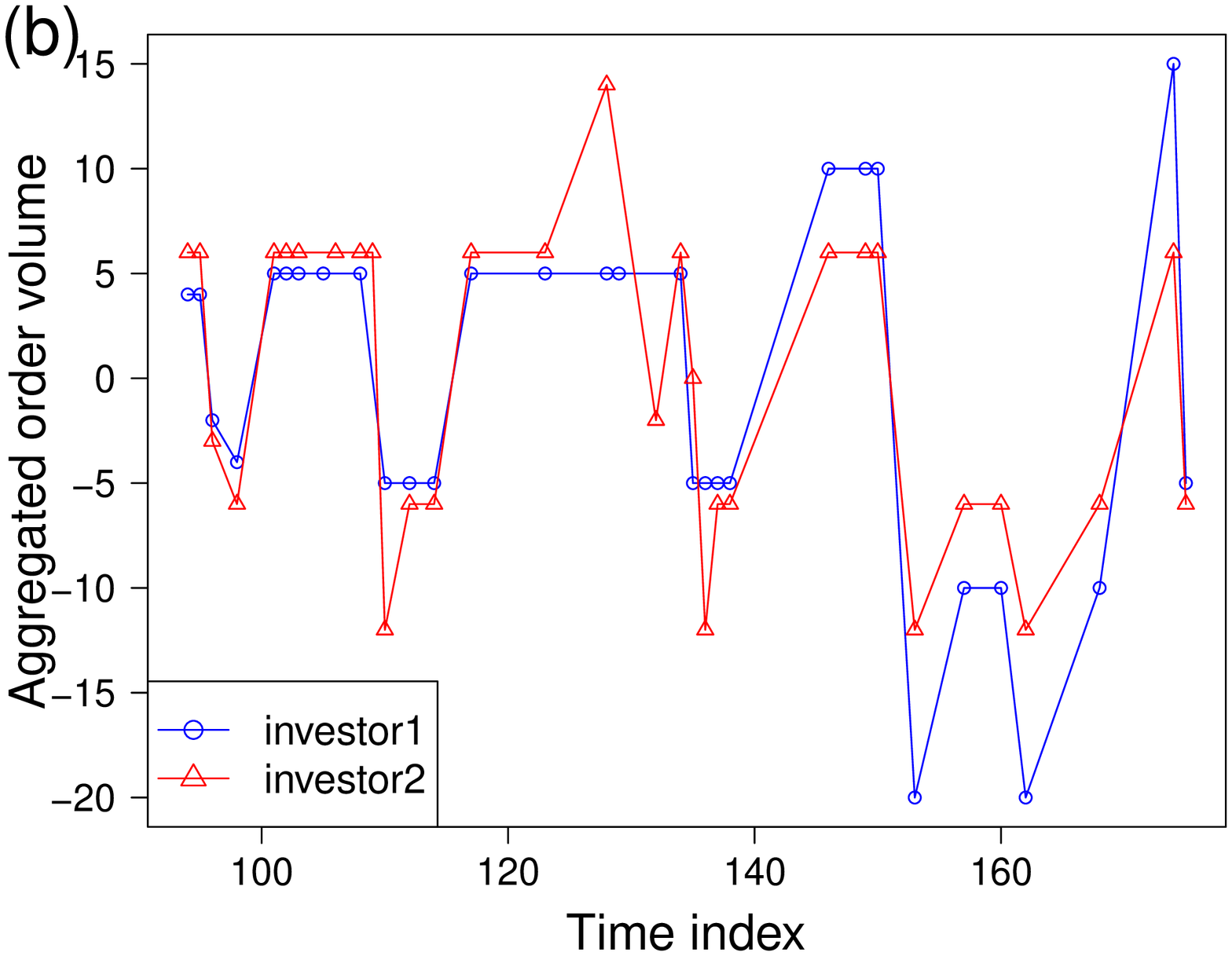}
    \caption{(a)~The time series of signed order volume of two investors and (b)~the corresponding aggregated time series with $\delta_t$=60 seconds. The aggregated time series with less data points retain the profile of the original time series.}
    \label{fig:time_series}
\end{figure}

\begin{figure}[tb]
    \centering
    \includegraphics[width=0.48\textwidth]{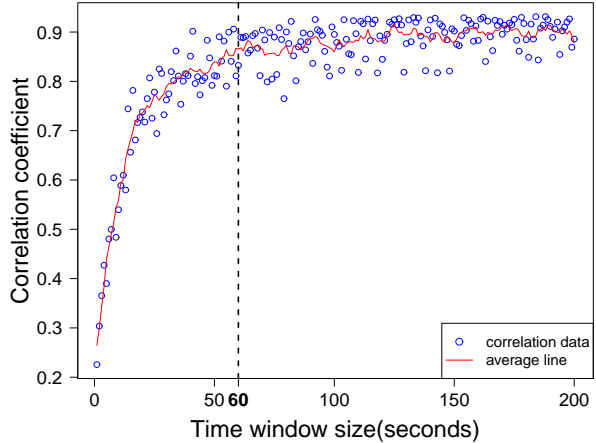}
    \caption{The impact of time window size $\delta_t$ on correlation coefficient when aggregating two time series of signed order volume.
    Each circle (in blue) indicates a correlation coefficient value at a certain window size. The average line (in red) can more evidently demonstrate the trend after smoothing data fluctuation. About 60 seconds are needed for the correlation
    coefficient to reach its asymptotically stable value, which means that it is reasonable to choose 60 seconds as the size of time window in our experiments.
    }
    \label{fig:window_size}
\end{figure}

In aggregating time series of signed order volume, the length of
time window $\delta_t$ is an important parameter that will directly
influence the correlation coefficient calculation. For examining the
impact of window size $\delta_t$ on correlation coefficient, we
choose two time series of signed order volume from the trading data
of fuel oil futures on September~25, 2008, which are shown
in~\autoref{fig:time_series}(a). We aggregate the two time series
with different sizes of time window. \autoref{fig:time_series}(b)
shows the aggregated time series with the time window size
$\delta_t$=60 seconds. Then we calculate the correlation coefficient
between the two resulting aggregated time series with the time
window size increasing from 1 to 200 seconds. The results are shown
in~\autoref{fig:window_size}. We can see that the correlation
coefficient increases as the size of time window enlarges, and it
reaches asymptotically to a stable value at about 60 seconds. The
result is analogous to the Epps effect~\cite{Epps1979, Reno2003,
Toth2007, Toth2009a} that the stock return correlation decreases as
the sampling frequency of data increases. From
~\autoref{fig:window_size}, we argue that a time window of size 60
seconds is a reasonable choice in our experiments.

\subsection{Determining the Length Threshold ($\delta_L$) of Aggregated Time Series}
\begin{figure}[tb]
    \centering
    \includegraphics[width=0.48\textwidth]{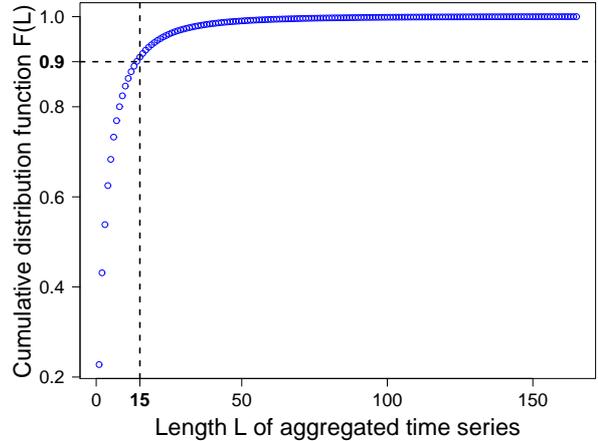}
    \caption{The cumulative distribution $F(L)$ of the length $L$ of aggregated time series over the whole data set.
    About 90\% time series are less than 15 (time windows) in length and are excluded from correlation coefficient calculation.}
    \label{fig:length_dist}
\end{figure}

For the whole data set, the cumulative distribution function
$F(L)$=$P(L'<L)$ of the length $L$ of aggregated time series is
shown in \autoref{fig:length_dist}. As the figure shows, about 90\%
time series are less than 15 in length and are excluded from
correlation coefficient calculation. There are only 10\% investors
included in collusive detection, which reduces the complexity of
correlation calculation. Therefore, we choose 15 as the empirical
threshold~($\delta_L$) value for filtering the short aggregated time
series, which means that an investor should have placed orders in at
least 15 time windows in a trading day to be included in the
collusive clique detection procedure. This choice conforms to the
long-term surveillance practical experience in the exchange
institute.

\subsection{The Effect of The Correlation Coefficient Threshold~$\delta_w$}
Now, we consider the order record data of the copper futures
contract in one typical trading day (September~18, 2008) to
demonstrate the process of collusive clique detection. After
aggregating and filtering the time series of signed order volume, we
obtain 819 eligible aggregated time series for computing the
correlation coefficient matrix $M_{c}$.

\begin{figure*}[tb]
    \centering
    \setlength\fboxsep{0pt}
    \setlength\fboxrule{0.5pt}
    \fbox{\includegraphics[width=0.48\textwidth,height=0.373\textwidth]{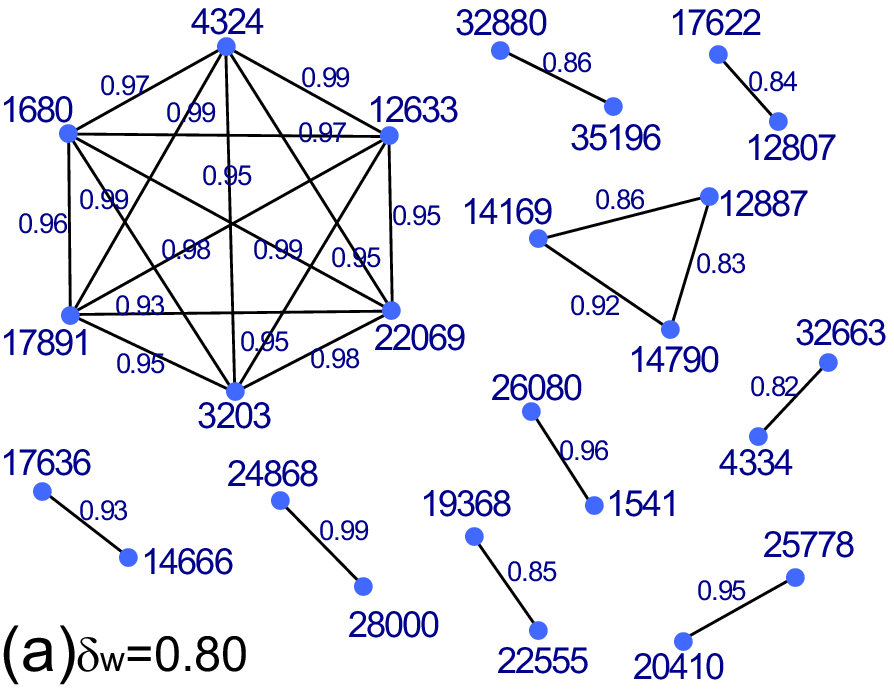}}
    \fbox{\includegraphics[width=0.48\textwidth,height=0.373\textwidth]{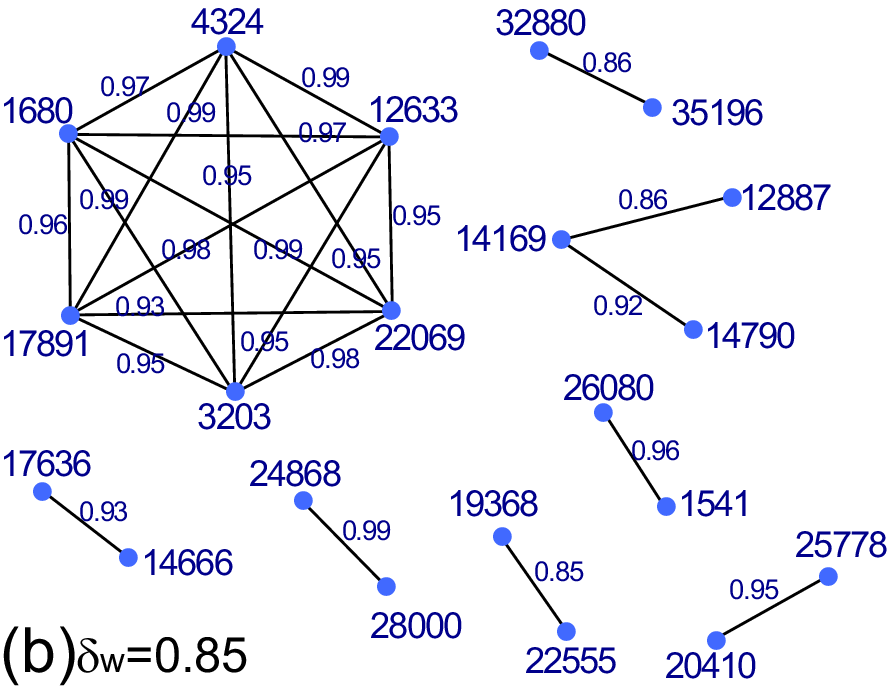}}
    \fbox{\includegraphics[width=0.48\textwidth,height=0.373\textwidth]{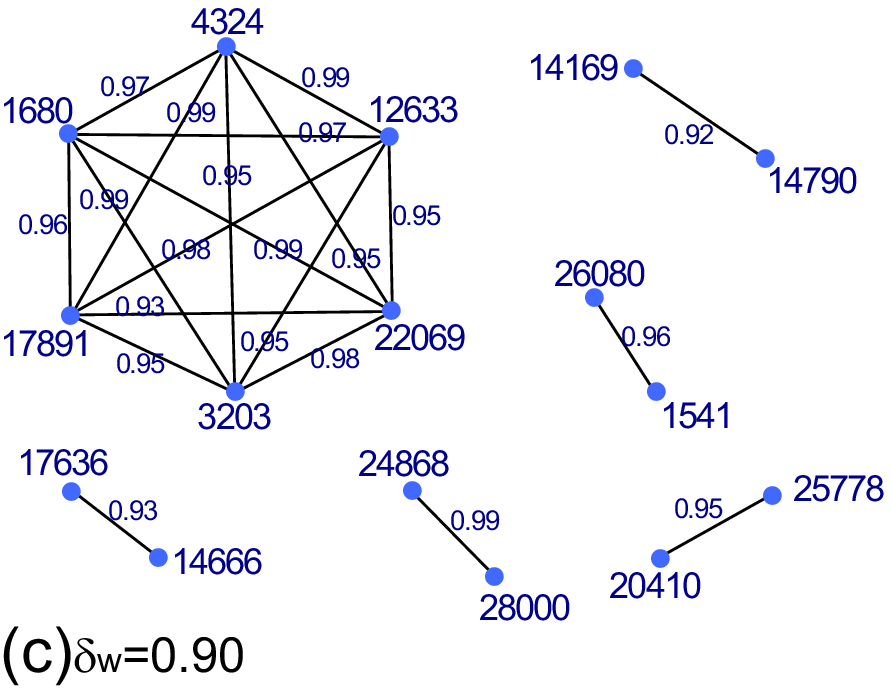}}
    \fbox{\includegraphics[width=0.48\textwidth,height=0.373\textwidth]{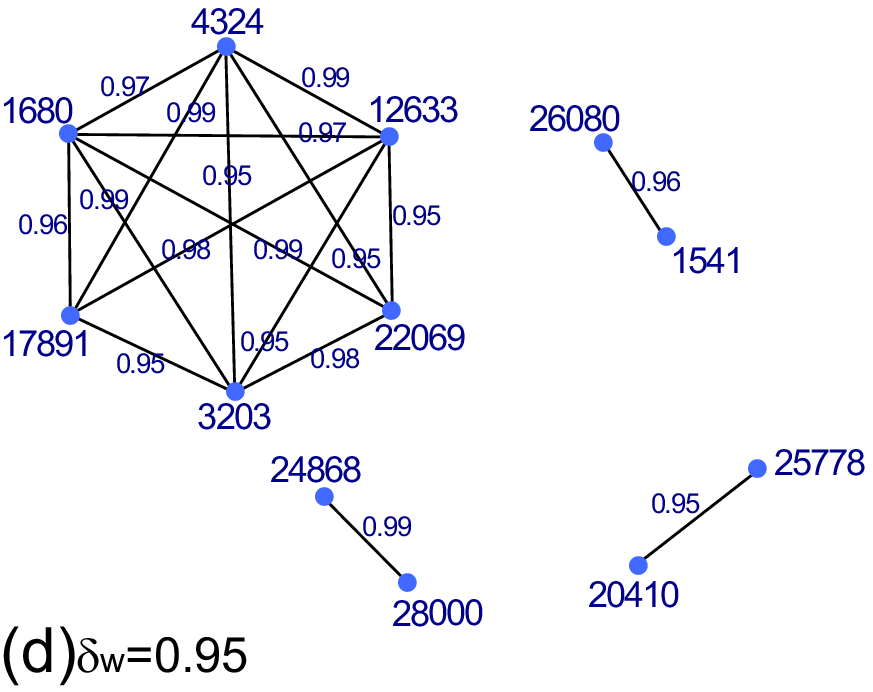}}
    \caption{The weighted graphs obtained by using different threshold $\delta_w$ values for the copper futures contract in a trading day. The number of resulting connected components reduces as the threshold value increases. The threshold value is adjusted for different monitoring levels in practical surveillance requirements.}     \label{fig:threshold_graphs}
\end{figure*}

We construct four weighted graphs based on $M_{c}$ with different
correlation coefficient threshold values. In
~\autoref{fig:threshold_graphs}, the number of connected
components are 10, 8, 6 and 4, corresponding to the threshold values
0.80, 0.85, 0.90 and 0.95, respectively. The number of resulting
connected components gradually decreases as the threshold value
grows. We notice that the connected component with six nodes is
(almost) a complete graph in all the sub-figures. The reason is that
the similarity between any pair of nodes in the component is very
large. In practical applications, the supervisors of the exchange
institute can choose different threshold values according to real
surveillance requirements to observe the suspect investors in
different monitoring levels.

\subsection{The Performance of the Proposed Method}
\begin{figure*}[tb]
    \centering
    \setlength\fboxsep{0pt}
    \setlength\fboxrule{0.5pt}
    \fbox{\includegraphics[width=0.48\textwidth,height=0.373\textwidth]{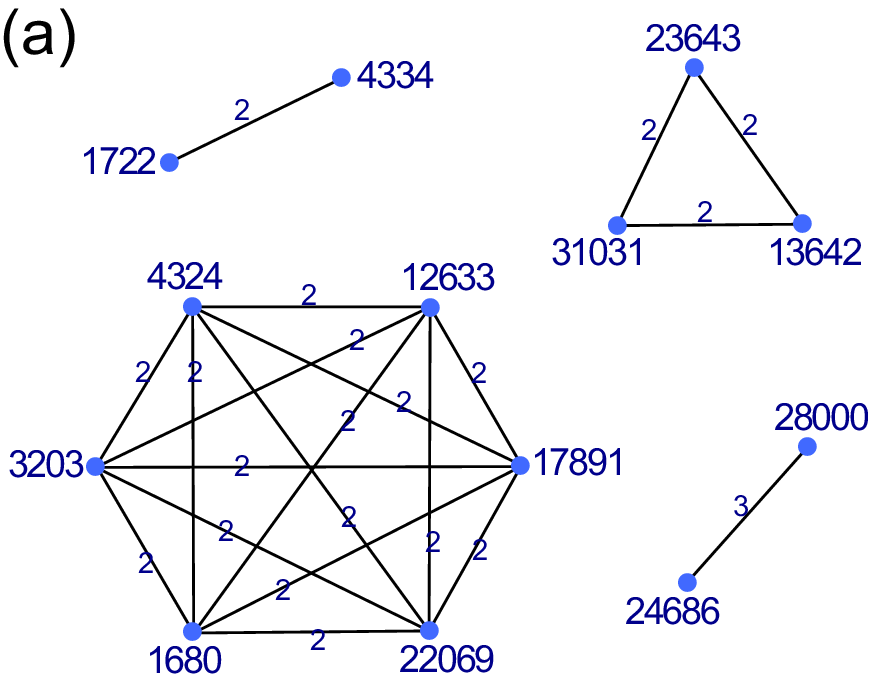}}
    \fbox{\includegraphics[width=0.48\textwidth,height=0.373\textwidth]{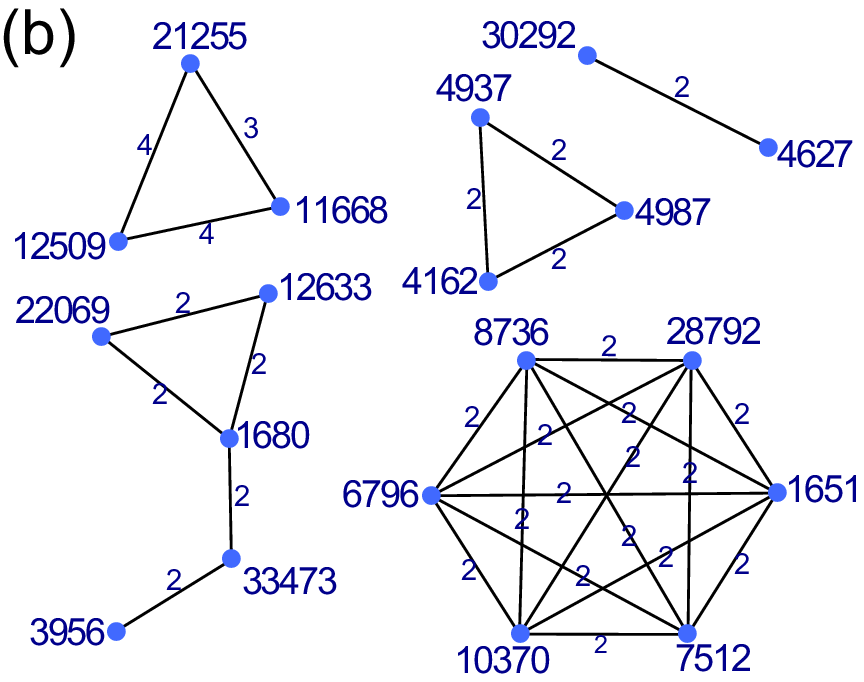}}
    \fbox{\includegraphics[width=0.96\textwidth,height=0.48\textwidth]{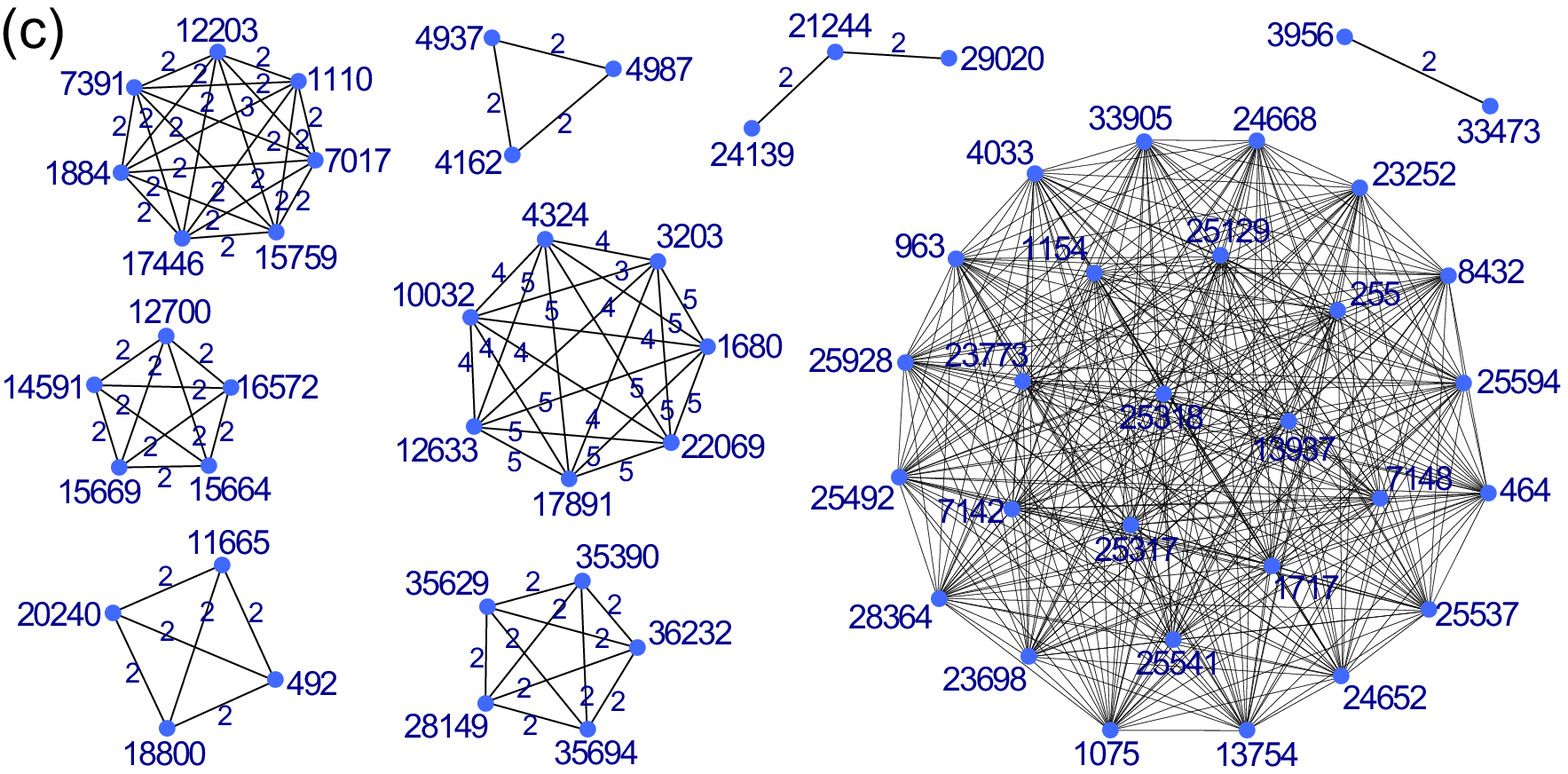}}
    \caption{The integrated weighted graphs of the copper~(a), fuel oil~(b) and  natural rubber~(c) futures by combining connected components of daily weighted graphs in nine consecutive trading days. The weight of each edge is the sum of its occurrences in each daily weighted graphs. Only those edges with weight no less than 2 are
    survived. Eventually, four (for copper), five (for fuel oil) and nine (for natural rubber) connected subgraphs are obtained, which will be output as suspect collusive cliques by our method. For the largest connected component at the bottom-right part of the figure (c), the edge weights are not shown due to space limit on the figure. However, we have computed the average value of its edge weights, which is 3.28.}
    \label{fig:results_graph}
\end{figure*}

According to the experimental results above, we choose the following
parameter values for the detection method: $\delta_t$=60 seconds,
$\delta_L$=15, $\delta_w$=0.90 and $\delta_f$=2. We construct the
daily weighted graphs of the three futures contracts
(see~\autoref{tab:odssi}) in nine consecutive trading days, and
merge the connected components occurring at least twice in the daily
graphs into an integrated weighted graph. We illustrate the
integrated weighted graphs for the copper, fuel oil and natural
rubber futures contracts respectively
in~\autoref{fig:results_graph}(a), \autoref{fig:results_graph}(b)
and \autoref{fig:results_graph}(c). There are eighteen connected
subgraphs in these figures. We can see that all the subgraphs are
complete graphs except the two ones \{22069, 12633, 1680, 33473,
3956\} in~\autoref{fig:results_graph}(b) and \{24139, 21244, 29020\}
in~\autoref{fig:results_graph}(c), and most subgraphs just appear twice
in the nine trading days, while the four subgraphs including
\{24686, 28000\} in~\autoref{fig:results_graph}(a), \{12509, 21255,
11668\} in~\autoref{fig:results_graph}(b), \{1680, 3203, 4324,
10032, 12633, 17891, 22069\} and the largest component
in~\autoref{fig:results_graph}(c) occur at least three times. This means
that these subgraphs can be considered as suspect collusive cliques.
The four subgraphs occurring more than three times can be more
confidently regarded as collusive cliques.

Furthermore, by carefully checking the figures, we notice that the
set of investors \{1680, 12633, 22069, 4324, 3203, 7891\}
forming a connected subgraph in~\autoref{fig:results_graph}(a)
and~\autoref{fig:results_graph}(c), and part of it \{1680, 12633,
22069\} appears in~\autoref{fig:results_graph}(b). In addition, the
two sets of investor \{3956, 33473\} and \{4162, 4937, 4987\} appear
in~\autoref{fig:results_graph}(b) and~\autoref{fig:results_graph}(c),
and the two sets of investors \{3956, 33473\} and \{1680, 12633,
22069\} are correlated in the fuel oil futures for they unify
together to a single subgraph in~\autoref{fig:results_graph}(b). So
we assert that these investor sets form collusive cliques with high
probability, which will be further confirmed by related background
data.

The experimental results for all the three futures contracts are
summarized in~\autoref{tab:exr}. The average number $N_a$ of
eligible aggregated time series in all the trading days are much
smaller than the number of corresponding investors
in~\autoref{tab:odssi}. This indicates that a large number of short
aggregated time series are excluded by the filter threshold
$\delta_L$ and only the active investors are kept for further
processing. In~\autoref{tab:exr}, there are many connected
components that occur only once in the nine trading days, though our
method will not classify them into suspect collusive cliques, the
exchange institute still needs to pay attention to them in the
following trading days. Certainly, these detected suspect collusive
cliques should be further probed and confirmed via the regulatory
procedure of the supervision system in the exchange institute. In
practice, these suspect cliques, even not confirmed, will be added
to the ``black" list of the surveillance system.


\begin{table}[tb]
    \captionstyle{indent}
    \caption{The experimental results of three futures contracts (including copper, fuel oil and natural rubber) in nine consecutive trading days.
    $\bar N_a$: the average number of eligible aggregated time series in the nine trading days;
    $N_c$: the number of the connected components in all weighted graphs;
    $N_s$: the number of the connected subgraphs in the integrated weighted graph, i.e., the number of detected suspect collusive cliques;
    $N_t$: the number of the verified collusive cliques by surveillance archival data of the exchange institute.
    }
    \label{tab:exr}
    \centering
    \begin{tabular*}{0.49\textwidth}{@{\extracolsep{\fill}} ccccc }
        \hline
        Futures & $\bar N_a$ & $N_c$ & $N_s$ & $N_t$ \\
        \hline
        \itshape copper         & 480   & 14 & 4 & 4 \\
        \itshape fuel oil       & 955   & 14 & 5 & 4 \\
        \itshape natural rubber & 1123  & 20 & 9 & 9 \\
        \hline
    \end{tabular*}
\end{table}

Up to now, we have found some suspect collusive cliques. Are they
really collusive cliques? or can we give some explanation on why
they are treated as collusive cliques. For this purpose, we firstly
examine the detected suspect collusive cliques carefully against
the surveillance archival data of the exchange. The archival data
covers the background information of all investors and companies
involved in the futures market. The findings are interesting and
promising. For most suspect collusive cliques, their members are
interrelated in one or another way. They either come from the same
community of a city or belong to the same company, or even they are
from a family. We also find that the accounts of some cliques are
controlled and operated by a backstage manipulator. The
interrelation information implies great possibility to concert
trading actions of members in a clique. Now we come to the final
step of this study: validate the detected suspect collusive cliques in
terms of verified collusive cliques of the surveillance system and
judgement of experienced domain experts from the exchange institute.
There are seventeen suspect collusive cliques verified as collusive
cliques. The numbers of verified collusive cliques(the column $N_t$
in~\autoref{tab:exr}) are 4, 4 and 9 for the futures contract copper,
fuel oil and natural rubber, respectively. Furthermore, we tracked and
analyzed the order records of the members of these cliques, and the
verified results were reconfirmed. The only detected suspect
collusive clique that is not verified is from the fuel oil futures
contract. The reason is that we can not find enough evidence. For
privacy reason, we can not provide any more detail of these detected
cliques.

It is worthy of pointing out that the detected suspect collusive
cliques or even the verified collusive cliques are not equal to
real collusive criminal cliques in financial markets. Nevertheless,
the detection results are still valuable to the market supervision
department as they can provide informative targets to the
supervision department for further probing, which is better than to
search potential financial criminal cliques from numerous investors
and massive trading data without any target.

\section{Conclusion and discussion}\label{sec:conclusion}
A method for detecting collusive cliques in futures trading markets
has been proposed under the framework of correlation analysis of
traders' behaviors and graphs merging. The proposed method defines
the aggregated time series to summarize signed order volume series
to achieve robust results, and then calculates correlation
coefficient matrix over all eligible unified aggregated time series
of signed order volume to construct weighted graphs, finally merges
the connected components from multiple weighted graphs corresponding
to multiple trading days into an integrated weighted graph.
Experiments are conducted to determine reasonable values for
different parameters, including the size of time window and other
thresholds, and detect suspect collusive cliques from real order
data of three futures contracts from the Shanghai Futures Exchange.
The major innovation of the proposed method lies in two aspects:
a)~the aggregated time series used to summarize signed order volume
series to alleviate the impacts of timestamp difference between
different order series and data fluctuation, and b)~the effective
scheme to compute the correlation coefficient between two
unevenly-spaced time series from irregular events. The proposed
method can also be applied to investigating other behavior
similarity of investors, for example, position changes per trading
day. Experimental results validate the effectiveness of the proposed
method. As a pilot application, a tool based on the proposed method
has been deployed in the Shanghai Futures Exchange, to assist
futures market surveillant and risk management.

As for future work, we are considering to further optimize the
method by utilizing the data of two neighboring time windows for
balancing the uneven data distribution. We also plan to take into
account more trading information such as canceled orders and trade
reports to enforce the information for detecting collusive.
Furthermore, we will explore effective approaches to detecting
collusive cliques that show ``different" trading
behaviors.

\section*{Acknowledgements}
We thank the anonymous reviewers for their valuable comments and suggestions. This work was supported by National Natural Science Foundation of China~(NSFC) under grants No.~60873040 and No.~60873070. Jihong Guan was also supported by the ``Shuguang" Scholar Program of Shanghai Education Development Foundation.

\section*{References}

\bibliographystyle{elsarticle-num}
\bibliography{References}

\end{document}